# THE FEASIBILITY OF ELECTRIC AIR TAXIS: BALANCING TIME SAVINGS AND CO$_2$ EMISSIONS - A JOINT CASE STUDY OF RESPECTIVE PLANS IN PARIS

N.Hagag[1] (https://orcid.org/0000-0001-7958-3322),
Bastian Hoeveler[2],

[1]DLR Institute of Flight Guidance,
German Aerospace Center
Lilienthalplatz 7, 38108 Braunschweig, Germany
Contact: nabil.hagag@dlr.de

[2]Axalp Technologies AG
Louis-Giroud-Str 26, 4600 Olten, Switzerland

**Abstract**

This paper presents a comprehensive evaluation of the sustainability of Advanced Air Mobility (AAM) within urban and regional mobility infrastructure, utilizing Paris as a prominent case study. Driven by ambitious environmental targets, Paris aims to transform its transportation landscape into a cleaner, safer ecosystem. Collaborating with public and private stakeholders, the region has positioned AAM as a promising facet of future mobility, highlighted by the world's first scheduled commercial electric Vertical Take-Off and Landing (eVTOL) air taxi service during the 2024 Olympic Games. The study's main goal is to assess the energy consumption and CO$_2$ emissions of AAM aircraft across typical flight missions, encompassing urban and regional routes. A comparison is drawn between eVTOL performance and conventional modes such as cars, public transport, and helicopters. Key findings reveal intriguing insights. On urban routes spanning 50 km, eVTOLs offer noteworthy time savings of around 23 minutes compared to cars and 22 minutes compared to public transport. Moreover, concerning specific scenarios, eVTOLs demonstrate substantial time savings for regional routes of 300 km—averaging 76 minutes compared to cars and 69 minutes compared to trains. Regarding CO$_2$ emissions, a contrast emerges between urban and regional contexts. Urban eVTOL operations are relatively less eco-friendly due to higher energy consumption, than electric cars. While multicopters consume 47% less CO$_2$ than traditional helicopters, they surpass petrol cars by 13%, diesel cars by 19%, and electric cars by up to 256%. In contrast, for regional travel, Lift-and-Cruise 1 eVTOLs consume 77% less CO$_2$ than average helicopters, 46% less than petrol cars, 44% less than diesel cars, but emit 68% more than electric vehicles and 96% more than electric trains. In summary, eVTOLs exhibit significant time savings and CO$_2$ reductions on regional routes, yet their overall environmental performance hinges on mission specifics. To harness AAM's full potential for Paris's sustainability goals, policymakers, manufacturers, and researchers should explore diverse configurations, account for real-world operations, and seamlessly integrate eVTOLs into the broader transportation framework. This approach can pave the way for greener, more efficient urban and regional transportation futures.

**NOMENCLATURE**

| | |
|---|---|
| $A_r$ | Rotor area |
| BM | Battery mass |
| $e_A$ | Effective energy density |
| $E_B$ | Provided energy of the battery |
| $E_C$ | Energy demand during cruise flight |
| $E_D$ | Energy demand of ground vehicle |
| $E_H$ | Energy demand during hovering flight |
| $f_c$ | Fuel consumption |
| g | Acceleration due to gravity |
| $h_v$ | Heat value |
| s | Distance |
| $\varepsilon_C$ | Lift-to-drag ratio |
| $\eta_C$ | Travel efficiency |
| $\eta_H$ | Hovering efficiency |
| $\mu_A$ | Weight ratio of BM and Max. Take-off Weight |
| ρ | Air density |
| AAM | Advanced Air Mobility |
| GHG | Greenhouse gas |
| PAX | Passenger |
| RAM | Regional Air Mobility |
| UAM | Urban Air Mobility |

## 1. INTRODUCTION

Advanced Air Mobility (AAM) is an air transport system concept that integrates new, transformational aircraft designs and flight technologies into existing and modified airspace operations [1]. Especially electric Vertical Take-off and Landing (eVTOL) vehicles are the focus of this new transport technology. Considering the growing sustainability awareness of potential customers, eVTOL concepts (e.g. air taxis) are advertised to be especially free of emissions and contributing to the reduction of greenhouse gas (GHG) emissions, while quiet enough to operate in urban or regional environments without disturbing residents.

The Paris Climate Action Plan, launched in 2018 outlines a comprehensive strategy to reduce GHG emissions by improving energy efficiency from buildings, transportation, and waste management. Based on this, the municipal of Paris wants to be a carbon neutral city, powered completely by renewable energy until 2050 [2].

Paris will host the Olympic Games 2024 and the world eagerly awaits the possibility of witnessing the first-ever commercial air taxi flight during this prestigious event [3]. This groundbreaking moment would not only enhance accessibility for travelers but also showcase the aviation industry's commitment to sustainability and time-saving convenience.

Additionally, Paris is building a new metro line that will provide a direct connection from the city to the airport by 2030 [4]. This crucial infrastructure improvement will greatly simplify the accessibility of the citizens and tourists to the airport and introduces another convenient and time-efficient transportation option. Paris clear focus is on innovative and sustainable transport solutions. [2]

However, are eVTOLs as sustainable in terms of time-efficiency as assumed, especially additionally considering their energy demand? The deployment of AAM in Paris raises questions about its sustainability, particularly in terms of energy demand and resulting $CO_2$ emissions. According to the International Energy Agency, aviation is responsible for approximately 2.5% of global $CO_2$ emissions [5]. While eVTOL aircraft may offer a more sustainable alternative to traditional helicopters, their energy demand and carbon footprint during operation must still be considered in its assessment.

The Paris Region is home to 18.3% of the French population with around 12.3 million inhabitants and is a gateway to Europe and the world. It is easy to access with three international airports and seven TGV high-speed train stations that connect it to all of the world's major economic centers [6]. Paris Region accounts for 70% of French train traffic, with five million passengers traveling by train in France every day, including 3.5 million in the Paris Region. The Gare du Nord, one of the ten main train stations in Paris, is the busiest station in Europe, with over 200 million passengers per year [6].

In the Paris region, the average number of daily trips per person is 3.8, but this number hides strong disparities. Parisians themselves travel the most with an average of 4.3 trips per day, but cover the shortest distance by 12 km. Conversely, residents of the outer suburbs travel farther by 24 km. It is the working population who travels the most, with an average of 4.3 trips per day. However, the daily time budget for travel is the same regardless of location, at 1h30 per day. Even in the outer suburbs, some people only travel within their local area, which balances the time budgets of those who go to Paris [7].

### 1.1. AIM

In this paper we address the question if eVTOLs can be a sustainable solution for urban or regional transportation in the Paris region. Specifically, the study will focus on the time saving, but also on the energy demand and carbon footprint of eVTOL aircraft during operation by conducting a joint case study of respective plans for AAM in Paris. In this regard, there are two essential aspects which contribute to the success of the air taxi technology:

*1) Do air taxis reduce commute travel time compared to conventional transportation solutions?*

*2) Do electric air taxis decrease the carbon footprint of travel compared to conventional transportation solutions?*

### 1.2. RELATED WORK

As part of DLR's internal project, HorizonUAM, the focus lies on evaluating the potential and challenges presented by air taxis and Urban Air Mobility (UAM) concepts. A study conducted within this project utilizes a drone traffic scenario generator and 4D trajectory planning technology, tested within the urban landscape of Hamburg, Germany. Through a comparative analysis of travel times and distances, the research underscores a noteworthy 50% reduction in travel time and an impressive up to 16% decrease in route length for air taxis compared to conventional taxicabs. [8]

The ASSURED-UAM project, led by Łukasiewicz Research Network – Institute of Aviation, represents a significant initiative to seamlessly integrate UAM with Air Traffic Management and urban transportation systems. This integration seeks to uphold UAM's acceptability, safety, and sustainability, thereby providing a valuable reference point. Through the analysis of energy efficiency parameters, offers a framework for contrasting UAM passenger transportation with conventional ground-based methods. This approach draws from insights obtained from urban mobility evaluations, facilitating a comprehensive assessment of UAM's energy efficiency within the broader transportation landscape. Addressing environmental ramifications, discernible trends emerge. Notably, smaller aircraft with modest payloads exhibit lower carbon footprints, particularly during operations compared to larger UAS. The interplay between carbon emissions and a nation's electricity mix becomes evident, with a clear correlation between fossil fuel contribution and carbon footprint. Furthermore, nuanced insights emerge regarding aircraft specifications, operational concepts, and infrastructure, each influencing carbon footprints across different phases of an aircraft's lifecycle. [9]

In summary, the amalgamation of these studies underscores a comprehensive understanding of UAM's sustainability implications, as well as its potential energy efficiency benefits. These insights can be harnessed to critically evaluate electric air taxis' carbon emissions, particularly in relation to traditional modes of transportation such as helicopters, gasoline, diesel, hybrid, and electric cars, public transport, and electric trains. The collective research showcases the urgency and global significance of devising environmentally sound urban mobility solutions.

## 2. STATE OF ART

The following chapter provides an overview of the latest concepts and advancements in the operation of AAM. This chapter also aims to offer a comprehensive understanding of the existing battery technologies used in eVTOLs and their essential for an environmentally friendly aerial transportation.

### 2.1. ADVANCED AIR MOBILITY

AAM refers to the use of aircraft in urban and regional areas to address traffic congestion and enhance overall mobility [10]. With the advancements in technology and the integration of Artificial Intelligence, AAM is expected to become a reality in Europe within the next 3-5 years [11]. While the term encompasses a broader range of use cases, this paper primarily focuses on the passenger transport aspects of AAM.

In this context, two main use cases for AAM are discussed: Urban Air Mobility (UAM) and Regional Air Mobility (RAM), as depicted in Fig. 1. Each of these scenarios presents a unique opportunity to revolutionize urban and regional transportation and offer fast, efficient, and environment friendly alternatives to conventional commuting methods.

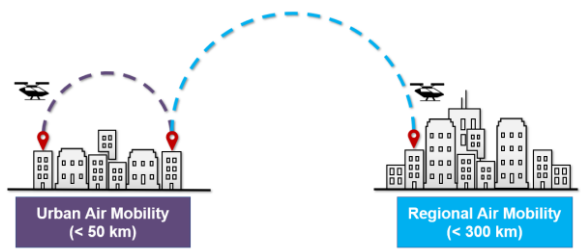

**FIG 1:** AAM based on UAM and RAM

Within the UAM scenario, eVTOL vehicles are designed to travel within a city, covering distances of about 50 km. This allows passengers to bypass long traffic jams and move swiftly to their destinations, contributing to smoother urban mobility.

On a larger scale, the RAM transport use case involves eVTOL vehicles traveling over distances of up to 300 km within highly urbanized areas. This holds the potential for valuable traffic relief and improved overall mobility in densely populated cities.

Each of these use cases presents its own set of challenges and opportunities that require careful evaluation and consideration to realize the full potential of AAM [10].

### 2.2. EVTOL TYPES

There are currently over 850 VTOL concepts with a large variety of technological maturity and different configurations [12]. Most of these concepts are propelled by electric motors supplied by battery systems. Some of them use a hybrid-electric approach where combustion engines act as generators. Essentially, the architectures used for these various concepts can be differentiated by the usage of a wing for high efficient cruise flight or being wingless [13]. In Fig. 2, the four most common types of eVTOL aircraft architectures are shown and compared against their forwarded and vertical lift. The choice of architecture already suggests the trade-off between the different cruise and hover efficiencies.

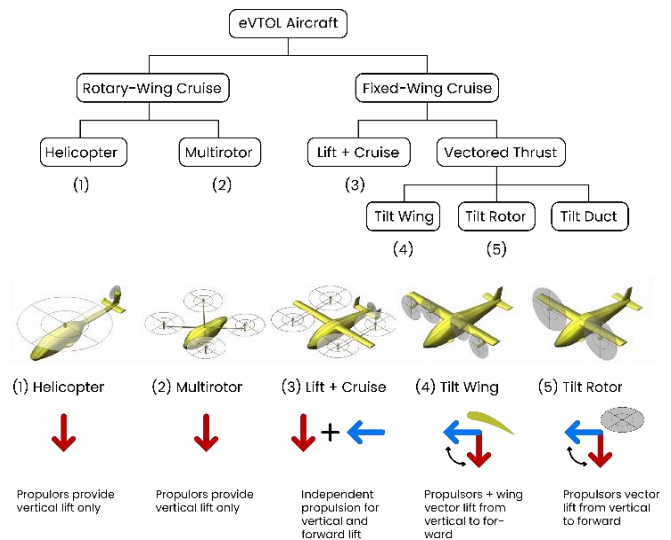

**FIG 2:** eVTOL configurations [14]

The multicopter configuration has high hover lift efficiency and low disc loading due to its high number of rotors. This means that it is able to take off and land vertically, but is less efficient during horizontal flight than other types due to high power demand.

Lift-and-cruise configuration have higher cruise efficiency and therefore able to fly longer distances compared to multicopters. This configuration is able to transit from vertical take-off to horizontal flight, allowing them to take advantage of both modes.

Tilt-rotor or tilt-wing configuration have lower hover lift efficiency and higher disc loading than multicopter or lift-and-cruise models, resulting in higher power demand and lower efficiency. However, these models are better suited for longer distances due to their ability to fly faster and their longer range.

Vectored-thrust or fixed wing configuration have low hover lift efficiency and high disc loading, meaning they are highly efficient in forward flight. However, they are less efficient in vertical take-off and landing than tilt-rotors or lift-and-cruise models.

## 2.3. LITHIUM-ION BATTERY

Batteries play a critical role in the operation of eVTOLs, as they provide the power required for the electric motors to lift the aircraft off the ground and maintain flight. The state of art in battery technology for eVTOLs is rapidly evolving, with research focused on increasing energy density, reducing weight, and improving safety. Lithium-ion batteries are commonly employed in the current generation of eVTOL aircraft due to their high energy density and well-established performance characteristics [12]. This battery type has up to today's state a gravimetric energy density between 150 - 350-Wh/kg and proven reliability.

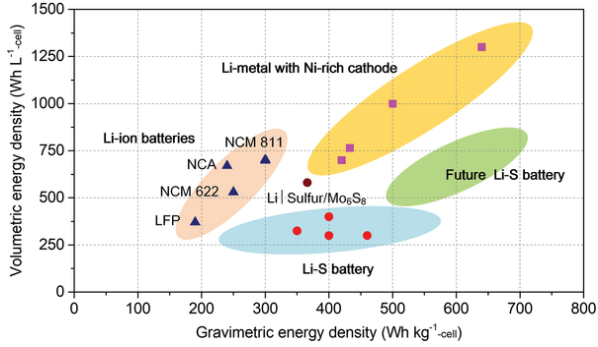

**FIG 1** Gravimetric & volumetric energy density of current battery development [15]

## 2.4. ENERGY DEMAND OF EVTOL

The energy demand of eVTOLs depend on various factors such as their weight, design, propulsion system, and operational mode. Generally, eVTOLs require considerable amounts of energy for the phase during VTOL compared to the horizontal flight. Additionally, is the energy demand further affected by the duration of hovering, but also weather conditions. Therefore, eVTOL manufacturers are continuously exploring ways to improve the efficiency of their vehicles through the use of lightweight materials, advanced propulsion systems, and optimized operational procedures. The energy demand of eVTOLs is a crucial factor in determining their commercial viability, as it directly affects operating costs, range, and environmental impact. [16, 17]

To determine the power and energy needed for hover flight of eVTOLs, it is necessary to calculate the thrust. [18] Based on the jet theory, thrust calculation equations can be derived, as described in [19]:

$$E_H = MTOW \cdot g \cdot \sqrt{\frac{MTOW \cdot g}{2\rho A_r}} \cdot \frac{1}{\eta_H} \cdot t_H \qquad (1)$$

Equation 2 determines the required power demand during cruise flight for all types of eVTOL. The formula for the required propulsion power during cruise flight differs from the power required in hover flight as it considers the glide ratio [19]:

$$E_C = MTOW \cdot g \cdot \frac{1}{\varepsilon_C} \cdot v_{real} \cdot \frac{1}{\eta_C} \cdot t_C \qquad (2)$$

The energy capacity of the battery depends on the total mass of the battery pack and the effective energy density:

$$E_B = e_A \cdot BM \text{ with } \mu_A = \frac{BM}{MTOW} \qquad (3)$$

## 2.5. ENERGY DEMAND OF GROUND VEHICLE

The energy demand for vehicles is a crucial factor in determining their environmental impact and overall efficiency. Petrol and diesel-powered vehicles have long dominated the automotive landscape, primarily relying on their internal combustion engines to generate power. The energy demand for these vehicles is closely linked to their fuel consumption, measured in litres per 100 kilometres (l/100 km).

In contrast, electric-powered vehicles, such as electric cars and trains, represent an eco-friendlier alternative. The energy demand for these vehicles is measured in kilowatt-hours per 100 kilometres (kWh/100 km). Electric vehicles rely on batteries to store electrical energy, which powers electric motors to drive the wheels. Their energy efficiency is considerably higher than traditional internal combustion engine vehicles, as they convert a larger portion of the energy from the grid into actual propulsion.

Using equation 4 to determine the required power demand during all types of ground vehicles, like cars and electric vehicles, but also public transportation by tram, bus or metro:

$$E_D = \frac{s}{100} * h_v * f_c \qquad (4)$$

| Vehicle | Ø $h_v$ [kWh/ltr] | Ø $f_c$ |
|---|---|---|
| Car – petrol | 8.2 | 8.4 ltr/100km |
| Car – diesel | 9.7 | 6.7 ltr/100km |
| Car – electric | - | 15 kWh/100 km |

**TAB 1** Average heat value and fuel consumption of ground vehicles

## 2.6. CARBON DIOXIDE EQUIVALENT / CO₂ (eq)

Based on Eurostat definition of carbon dioxide equivalent ($CO_2$ equivalent or $CO_2$ (eq)) is a metric measure used to compare the emissions from various greenhouse gases on the basis of their global-warming potential, by converting amounts of other gases to the equivalent amount of $CO_2$ with the same global warming potential. This potential is a measure for the atmospheric warming caused by a gas compared to $CO_2$, which has the factor 1 for $CO_2$. [20]

The carbon footprint of electric vehicles depends on the energy mix used to generate the electricity. If the electricity comes from renewable sources, such as wind or solar power, the carbon footprint of electric vehicles can be close to zero. The considerations of practical terms, certain emissions associated with the production, transportation, and installation of renewable energy infrastructure may be accounted for, thus affecting the total emissions over the lifecycle. However, if the electricity comes from power plants, such as coal-fired plants, the carbon footprint of electric vehicles remains substantial by around 50 grams of $CO_2$ per kilometre, which is still better than conventional vehicles. [25]

The $CO_2$ emissions depend on the energy demand and the energy source. Even fully electric UAM vehicles are not completely free of carbon emissions, as the source of the electricity used to charge the batteries has an essential impact on the carbon footprint of the vehicle. [24]

Recent studies have shown that the primary energy demand and $CO_2$ emissions of eVTOLs are notable lower than those of conventional aircraft. According to a report by Roland Berger, eVTOLs reduce $CO_2$ emissions by up to 50% compared to conventional helicopters [21]. Additionally, a study by the University of Michigan found that eVTOLs can reduce up to 40% energy demand compared to ground-based electric cars [22]. These results demonstrate the potential for eVTOLs to improve sustainability.

The study by Carnegie Mellon University examines the energy demand and GHG emissions of a very small quadcopter drone used for last-mile deliveries. The model showed that an electric quadcopter drone transporting a package off 0.5 kg consumes 0.08 MJ/km and causing 70 g of $CO_2$ (eq) considering the electric energy mix in the United States. Comparisons with other vehicles show that drones can reduce the energy consumption by 94% and 31% and GHG emissions by 84% and 29% per package delivered by replacing diesel trucks and electric vans, respectively [23].

To obtain realistic $CO_2$ emissions for an UAM operation based on the energy demand, flight scenarios are designed for the use case Paris. The following formula describes the calculation of $CO_2$ emission based on the hover and cruise energy demand by:

$$CO2\ emission = energy\ demand\ (E_S + E_R) * energy\ mix \quad (5)$$

## 3. METHODOLOGY

This chapter describes the approach to determine the $CO_2$ emissions of helicopters and UAM vehicles for relevant routes in context of the Paris 2024 Olympic games.

### 3.1. SELECTED FLIGHT MISSION PROFILE

In order to determine the energy demand of eVTOLs, the flight scenario is divided into two main phases, namely hover and cruise. The energy demand of the selected eVTOLs is evaluated by analysing the calculated results based on the selected properties under optimal conditions as shown in the following figure.

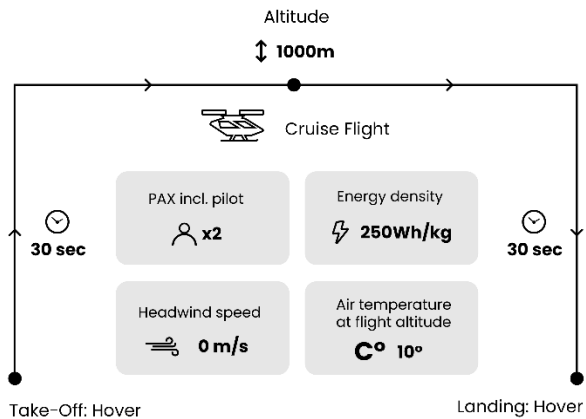

**FIG 5** UAM mission profile

The energy demand calculations are performed by integrating all equations and data into a macro calculation tool, based on previous research findings. The calculation tool allows for the selection of different eVTOL configurations and flight mission profiles.

### 3.2. SELECTED FLIGHT ROUTES

In this paper, the starting point is the Place de la Concorde in the city centre of Paris. According to the City of Paris, which provides real-time traffic updates through its website, traffic congestions are often high between 2 PM and 4 PM in the city centre.

By using Google Maps the average distance and travel time of seven days are calculated to ensure a representative analysis. This step is repeated for both the use cases UAM (max. 50 km) and RAM (max. 300 km).

The first set of missions contain UAM routes between places that are relevant for the Olympic games. They start at Place de la Concorde and go to 8 relevant points of interest in Paris with a distance between 2 km and 29 km.

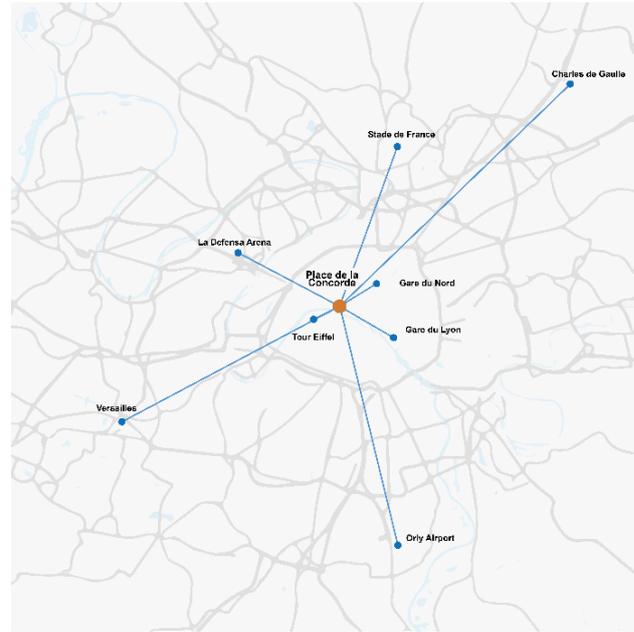

**FIG 6** UAM mission use case (<50 km)

| UAM (< 50 km) | |
|---|---|
| **Destination** | **Flight range [km]** |
| Eiffel Tower | 1,9 |
| Gare du Nord | 3,2 |
| Gare du Lyon | 4,6 |
| La Défense Arena | 7,4 |
| Stade de France | 7,5 |
| Orly Airport | 15 |
| Versailles | 16 |
| Charles de Gaulle Airport | 29 |

**TAB 3** Selected flight destinations on urban level

The second set of missions contain RAM routes between Place de la Concorde and 8 cities in the region of France with a distance between 69 km and 240 km.

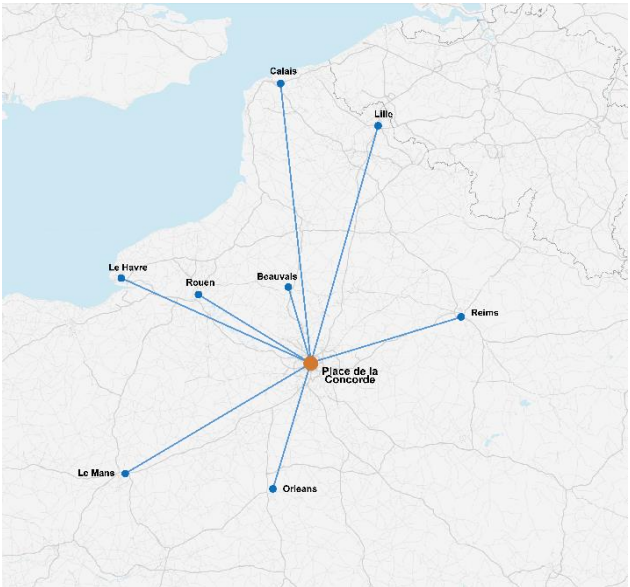

**FIG 7** RAM mission use case (<50 km)

| RAM (< 300 km) ||
|---|---|
| **Destination** | **Flight range [km]** |
| Beauvais | 69 |
| Rouen | 101 |
| Orleans | 103 |
| Reims | 140 |
| Le Havre | 179 |
| Le Mans | 185 |
| Lille | 205 |
| Calais | 240 |

**TAB 4** Selected flight destinations on regional level

All distances are determined under the assumption of a direct flight path without considering airspace structure, routing restrictions and topography influence.

### 3.3. SELECTED EVTOLS

The initial stage of the investigation involves determining the characteristics for various eVTOL vehicles.

| eVTOL type | Cruise speed [hm/h] | Battery mass [kg] | Hover efficiency $\eta_S$ | Cruise efficiency $\eta_R$ | Drag ratio $\varepsilon_R$ |
|---|---|---|---|---|---|
| Multicopter | 100 | 180 | 0,70 | 0,50 | 4 |
| Multicopter (coaxial) | 100 | 100 | 0,55 | 0,48 | 4 |
| Quadcopter | 120 | 800 | 0,55 | 0,40 | 2 |
| Lift+Cruise 1 | 180 | 200 | 0,70 | 0,50 | 9 |
| Lift+Cruise 2 | 180 | 400 | 0,70 | 0,50 | 11 |
| Vectored Lift | 300 | 900 | 0,60 | 0,60 | 17 |
| Tilt-rotor | 320 | 400 | 0,70 | 0,46 | 15 |

**TAB 5** eVTOL type configuration [24]

For the UAM use case with a range of up to 50 km, a multicopter is a suitable choice due to its inherent ability to take off and land vertically, enabling operations within constrained urban spaces. While other configurations such as vectored-thrust, lift-and-cruise, and tilt-rotor can also achieve VTOL capability, the use of a multicopter configuration may offer advantages in terms of maneuverability and adaptability to urban environments. However, these alternative configurations might present challenges related to their maneuverability within densely populated urban areas.

In contrast, for the RAM scenario, a lift-and-cruise model is a better choice. These models have the ability to take off and land vertically, but can then transition into a more efficient cruise mode, allowing them to cover longer distances at higher speeds. This makes them more suitable for longer, regional flights that require higher speeds and greater efficiency.

### 3.4. SELECTED HELICOPTERS

A total of seven helicopters have been assessed for their $CO_2$ emissions per passenger-kilometer during flight. The helicopters under investigation are detailed in Table 6. This range encompasses helicopters like the Robinson R44, accommodating 3 passengers and possessing a maximum take-off mass of 1089 kg, all the way up to the medium-sized Airbus Helicopter H145, with a capacity of up to 10 passengers and a maximum take-off mass of 3900 kg. With the exception of the Robinson R44, which utilizes gasoline, all helicopters employ kerosene as fuel. For reference, the combustion of 1 gallon of kerosene emits 9.9 kg of $CO_2$, while the same quantity of gasoline emits 8.8 kg of $CO_2$ [25].

The helicopters' hourly burn rate and cruise speed are detailed in their flight manuals. Just like with the UAMs, flight time for the helicopters is determined by dividing the direct distance by the cruise speed, without accounting for airspace layout, terrain, or approach and departure procedures. Furthermore, an additional allowance for factors such as run-up times, taxiing, route deviations during cruise, and approach and departure procedures is factored into the calculations.

| Helicopter Type | PAX seats | Cruise speed [km/h] |
|---|---|---|
| R44 | 3 | 200 |
| R66 | 4 | 200 |
| H120 | 4 | 223 |
| H125 | 6 | 260 |
| H135 | 7 | 253 |
| H145 | 10 | 247 |
| Bell 206 | 3 | 223 |

**TAB 6** Helicopter type configurations

### 3.5. SELECTED ENERGY MIX

The $CO_2$ emissions generated by the operation of air taxis are closely related to the electricity mix of the country in which the operation takes place. The data collected serves as the basis for determining energy demand and $CO_2$ emissions. In this regard, the average energy demand of the chosen mode of transport and the emission factor of the electricity mix of the average consumption in Europe are considered. These data are combined with suitable calculation methods to determine the associated $CO_2$ emissions.

| Area | Ø Energy mix [g CO$_2$/kWh] |
|---|---|
| Sweden | 13 |
| France | 57 |
| Germany | 508 |
| Europe | 226 |

**TAB 7** Energy mix by different regions [26]

Based on the energy demand calculated it is possible to calculate the CO$_2$ emissions during the operation for each use case.

### 3.6. SELECTED TRAVEL EMISSION CALCULATIONS

All emission values utilized in this study are derived from the "Travel Emissions Calculator" provided by GoClimate, which estimates carbon emissions generated by different travel methods based on distance [27]. The calculator then provides the estimated carbon emissions for each mode of transport such as petrol, diesel, or electric car, as well as train and subway. It is essential to note that the emission data used in the calculator is based on internal data, and therefore, the results may vary depending on the geographical location of the user.

| Vehicle type | Ø CO$_2$ emissions per PAX [CO$_2$ g/km] |
|---|---|
| Car (petrol) | 110 |
| Car (diesel) | 100 |
| Hybrid car | 65 |
| Electric car | 35 |
| Public transportation | 64 |
| Electric train | 36 |

**TAB 8** Travel CO$_2$ emission

The utilization of the GoClimate Travel Emissions Calculator allows us to obtain reliable and standardized emission values for various transportation modes, forming a solid foundation for evaluating the environmental impact of travel in our study.

## 4. RESULTS

This chapter presents the findings regarding time savings and CO$_2$ emissions associated with eVTOLs, focusing on the selected routes and mission profile within the context of Paris and France. As mentioned in Chapter 3.3, the analysis covered two representative eVTOL configurations: a multicopter for the UAM use case and a lift-and-cruise model for the RAM use case.

### 4.1. TIME SAVING

The results show that the average time savings is approximately 23 minutes when using a multicopter compared to a car, while compared to public transportation, the average time savings is 22 minutes. These results were obtained considering only the pure flight time and assuming direct flight paths for the selected urban destination in Paris. Furthermore, no boarding time was considered.

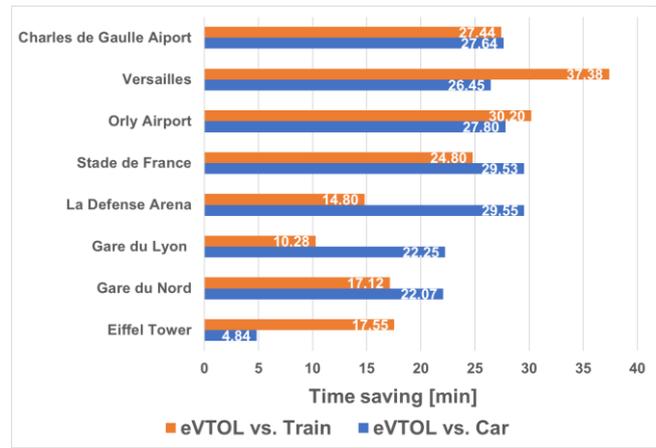

**FIG 8** UAM Time saving: eVTOL vs. Car vs. Train

On regional level, the diagram below shows the average time savings. When using a lift-and-cruise model, the average time savings is approximately 76 minutes compared to a car, while compared to public transportation, the average time savings is 69 minutes. These results were obtained considering only the pure flight time and assuming direct flight paths for the selected regional destinations in France departing from Paris. Again, no boarding time was considered.

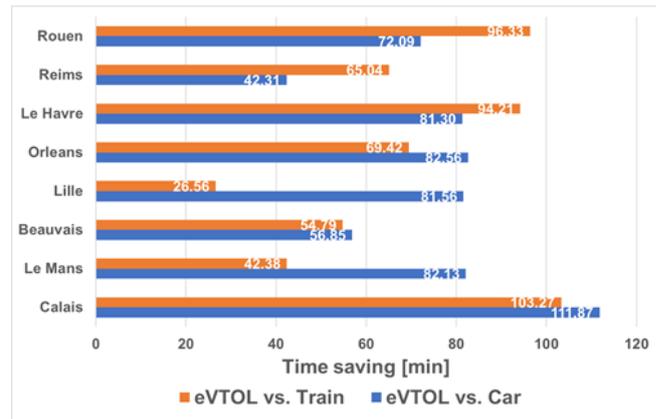

**FIG 9** RAM Time saving: eVTOL vs. Car vs. Train

### 4.2. CO$_2$ EMISSION MULTICOPTER VS. CARS

In this section, the results of the study on the comparison of CO$_2$ emissions originating from the energy demand of multicopter and conventional passenger cars (powered by petrol, diesel, hybrid, and electric) in an urban and regional setting are shown.

It is evident that multicopters for the use case UAM generally exhibit higher CO$_2$ emission values per passenger compared to conventional cars.

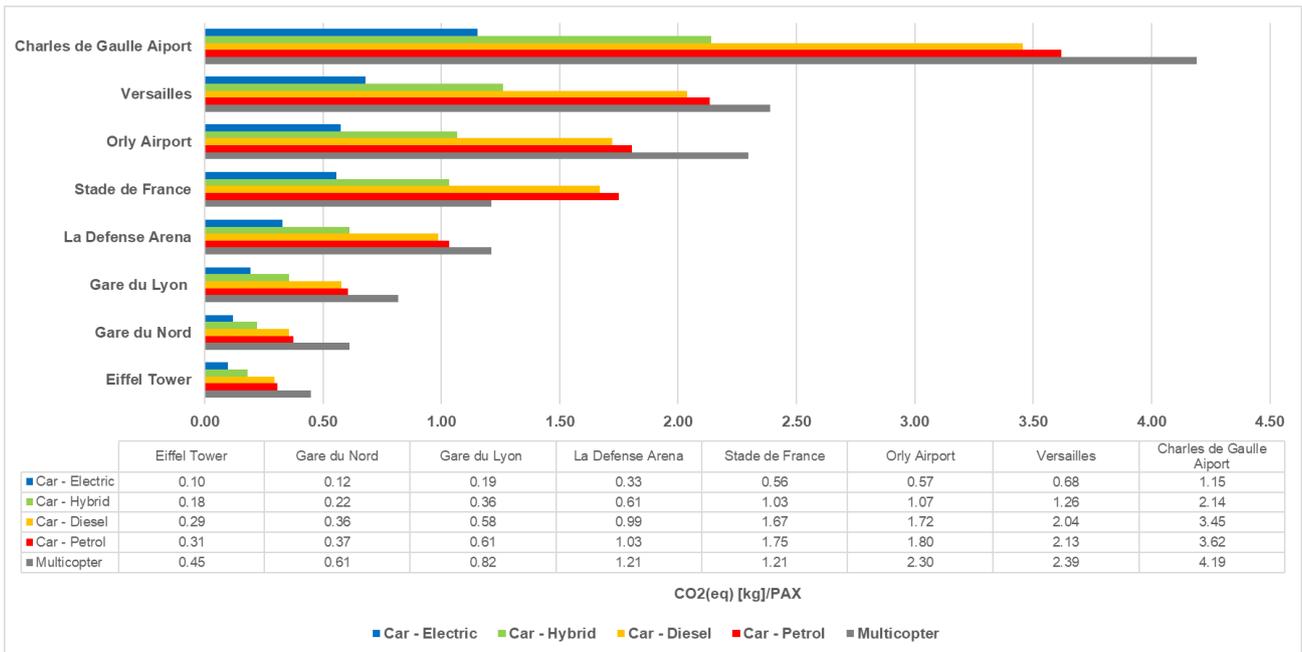

**FIG 10** Use Case: UAM – total $CO_2$ emission comparison multicopter vs. car

The highest emission value of 4.19 kg $CO_2$ (eq) by a multicopter is recorded for the distance from Place de la Concorde to Charles de Gaulle Airport, covering a distance of 28.94 km. Comparatively, the electric car requires only 1.15 kg $CO_2$ (eq) for this distance, the hybrid car 2.14 kg $CO_2$ (eq), the diesel car 3.45 kg $CO_2$ (eq), and the car by petrol 3.62 kg $CO_2$ (eq). Interestingly, on the shortest distance to the Eiffel Tower, spanning only 2.8 km, multicopter emit a relatively low $CO_2$ emission of 0.45 kg $CO_2$ (eq), suggesting their potential as a more environmentally friendly option for short-distance travel, compared to some conventional cars such as a patrol car with 0.31 kg $CO_2$ (eq), diesel car with 0.29 kg $CO_2$ (eq), hybrid car with 0.18 kg $CO_2$ (eq) and electric car with 0.10 kg $CO_2$ (eq). The electric vehicle records the lowest $CO_2$ emission consumption for this route.

The following figure illustrates the resulting percentage comparison of $CO_2$ emissions between multicopter and different types of car per passenger concerning their respective total emissions. It is evident that for the aforementioned example of traveling from Place de la Concorde to Charles de Gaulle Airport, using an eVTOL for the flight route results in approximately 16% higher $CO_2$ equivalent emissions compared to the driving route with a petrol-powered car. When compared to a diesel-powered car, the eVTOL emits around 21% more $CO_2$, approximately 96% more than a hybrid vehicle, and a striking 364% more than an electric vehicle. Regarding the shortest distance to the Eiffel Tower, similar trends are observed: The eVTOL produces about 46% more $CO_2$ equivalent emissions than a gasoline car, roughly 53% more than a diesel car, approximately 247% more than a hybrid vehicle, and a significant 459% more emissions compared to an all-electric car.

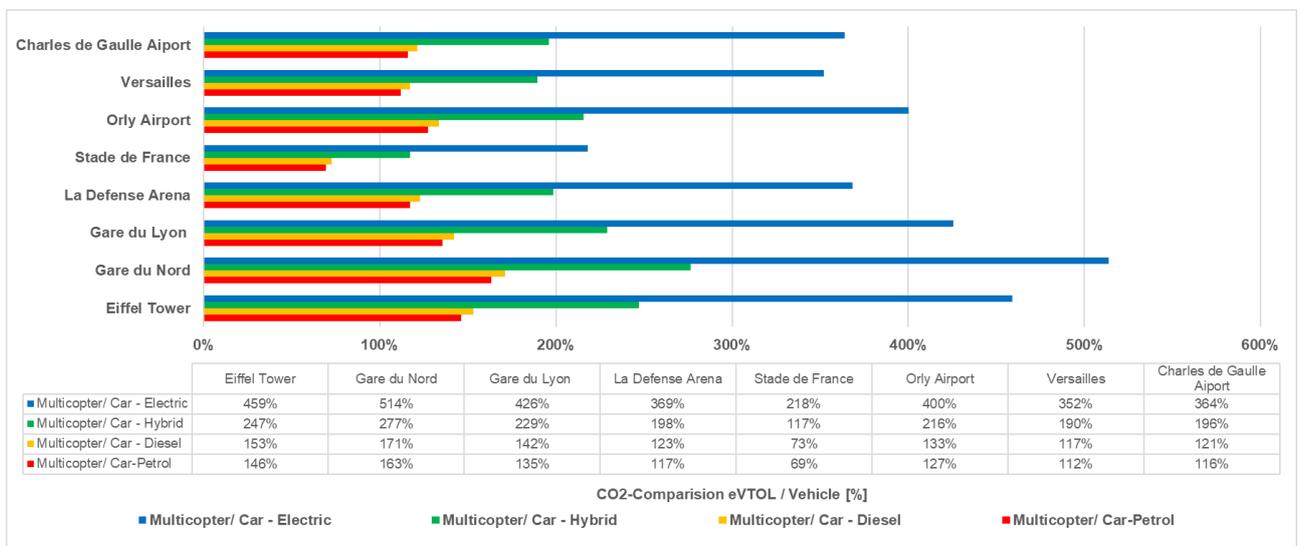

**FIG 11** Use Case: UAM – percentage CO2 emission comparison multicopter vs. car

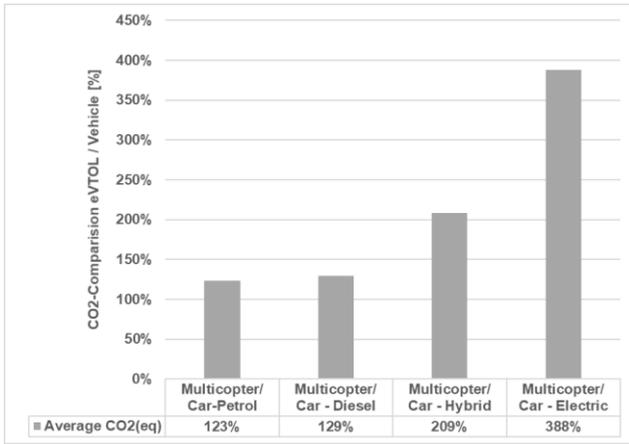

**FIG 12** Use Case: UAM – average percentage $CO_2$ emission comparison eVTOL vs. car

On average across all 8 flight missions, an multicopter emits approximately 123% more $CO_2$ equivalent emissions compared to a car running on petrol, roughly 129% more than a car powered by diesel, approximately 209% more than a hybrid vehicle, and a staggering 388% more than an electric car.

In relation to the regional use case (RAM), the results reveal that concerning the highest emission value of 16.82 kg $CO_2$ (eq) by the lift-and-cruise eVTOL is recorded for the distance from Place de la Concorde to Calais, covering a direct distance of 240.39 km. Comparatively, the electric car requires only 10.43 kg $CO_2$ (eq) for this distance, the hybrid car kg 19.37 $CO_2$ (eq), the diesel car 31.29 kg $CO_2$ (eq), and the car by petrol 32.78 kg $CO_2$ (eq).

The shortest distance is between Paris and the city of Beauvais with a direct flight distance of 69.46 km. For this route, the $CO_2$ emission per passenger is calculated at 5.12 kg $CO_2$ (eq). Comparatively, the electric car requires only 3.40 kg $CO_2$ (eq) for this distance, which accounts for 48% of the ratio. Compared to Car-Diesel the eVTOL configuration consumes nearly 50% less $CO_2$ emissions. Moreover, it results in 19% less emissions than a hybrid car however 51% more emissions than an electric car.

The following figures 13 and 14 presents also the total and percentage $CO_2$ emission comparison of $CO_2$ emissions between the lift-and-cruise configuration and different car types at regional level.

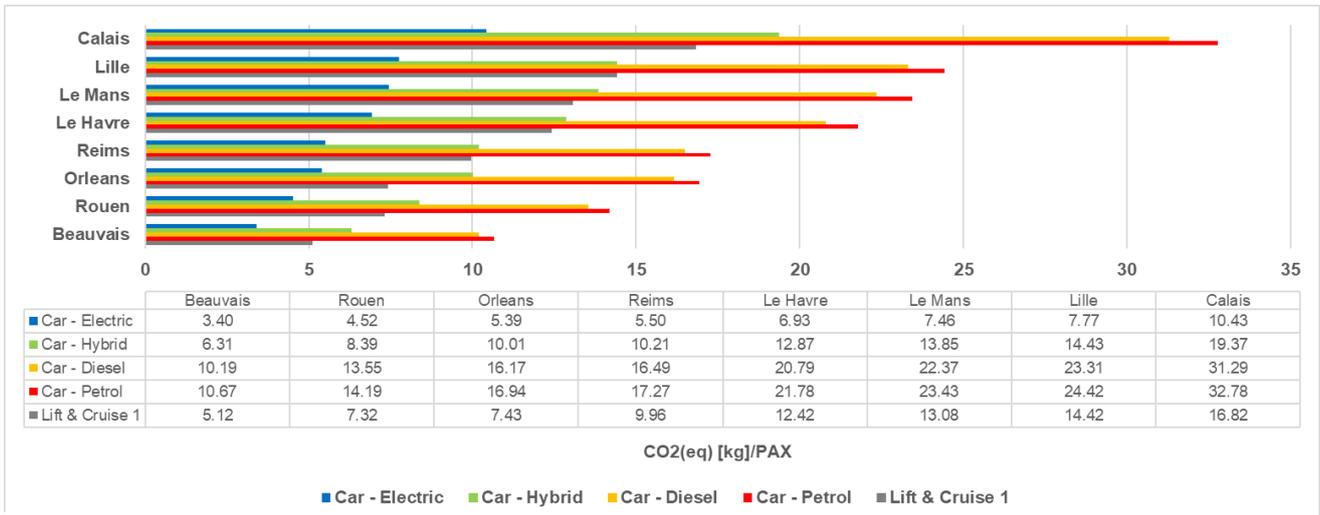

**FIG 13** Use Case: RAM – total CO2 emission comparison lift-and-cruise 1 vs. car

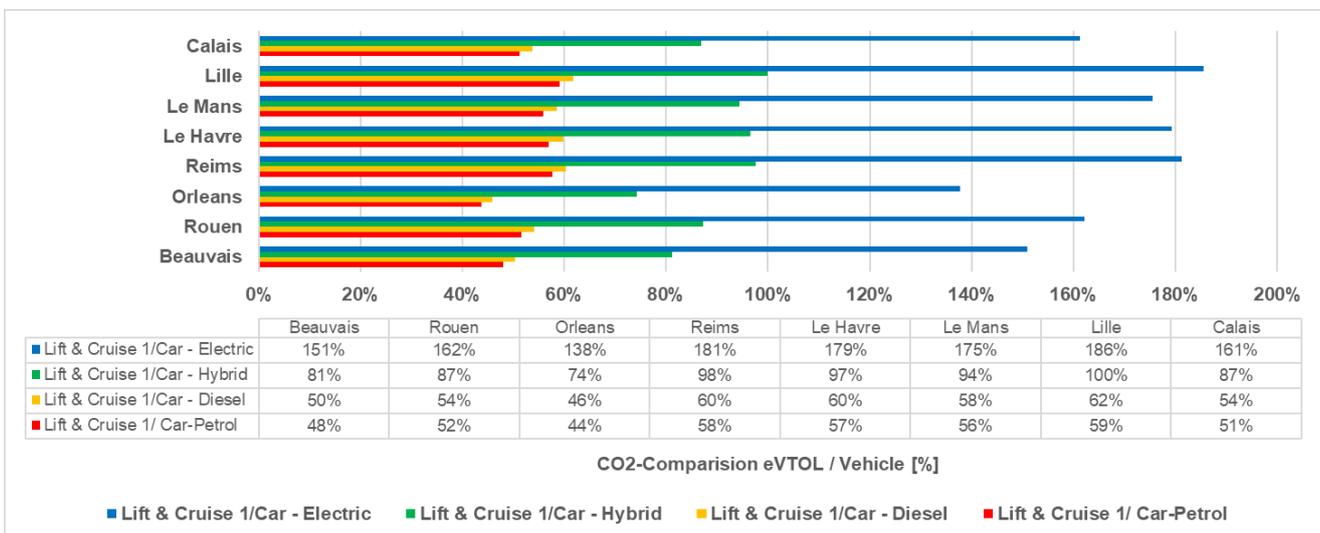

**FIG 14** Use Case: RAM – percentage $CO_2$ emission comparison lift-and-cruise 1 vs. car

On average across the 8 selected regional flight routes, an eVTOL can save around 47% of $CO_2$ emissions compared to a patrol car, approximately 44% compared to a diesel car, and approximately 10% compared to a hybrid car. However, when compared to electric vehicles, eVTOL emissions are approximately 67% higher.

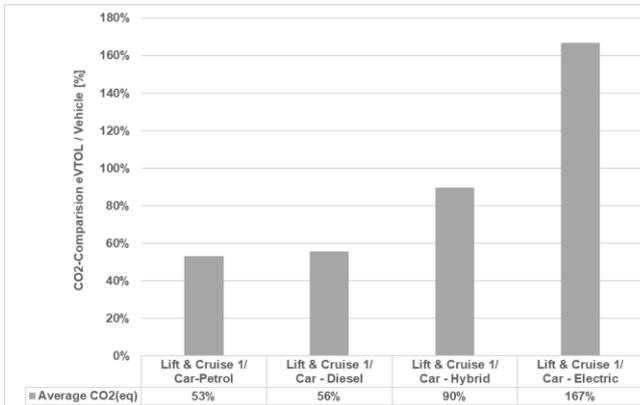

**FIG 15** Use Case: RAM – average percentage $CO_2$ emission comparison eVTOL vs. car

These findings indicate that, on a regional level, eVTOLs present notable advantages in reducing $CO_2$ emissions when compared to conventional petrol and diesel cars. Nevertheless, they still emit about notable more $CO_2$ than electric vehicles on average for the 8 selected regional flight routes.

## 4.3. CO2 EMISSION EVTOL VS. TRAIN

In this section, the results of the study on the comparison of $CO_2$ emissions from energy demand between eVTOLs and public transportation on urban and regional level are presented. Following figure shows the total $CO_2$ emissions at urban level between multicopter eVTOLs and public transportation (average value of metro, tram and bus, normalized to 2 PAX).

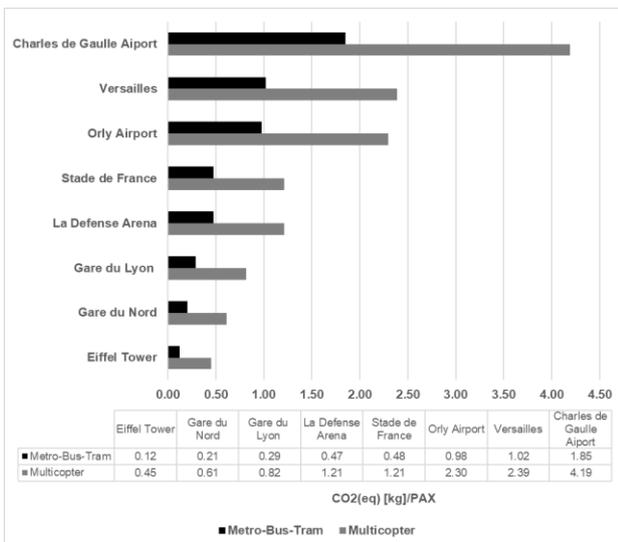

**FIG 16** Use Case: UAM – total $CO_2$ emission comparison multicopter vs. public transportation

From the quantitative results it is evident that the total $CO_2$ emissions generated by eVTOLs are crucial higher than those produced by public transportation.

The following figure presents a percentage comparison of $CO_2$ emissions between multicopter eVTOLs and public transportation.

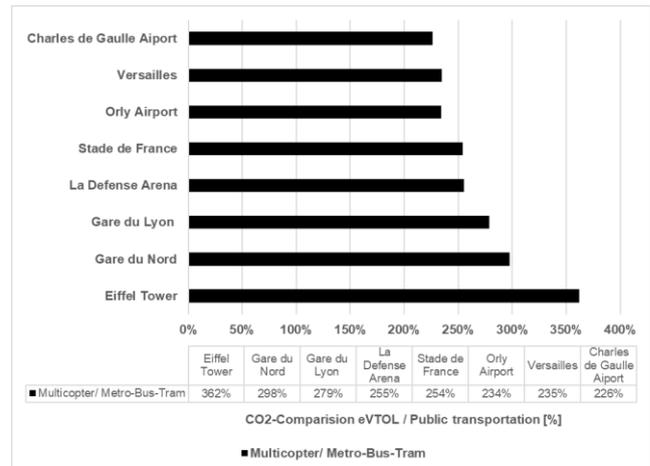

**FIG 17** Use Case: UAM – percentage $CO_2$ emission comparison multicopter vs. public transportation

The $CO_2$ emission results show that on urban level an eVTOL compared to public transportation, normalized to 2 passengers, exhibits crucial higher $CO_2$ emissions. For the flight route from Place de la Concorde to Charles de Gaulle Airport, eVTOLs produce approximately 226% more $CO_2$ emissions. Similarly, for the route to the Eiffel Tower, eVTOLs generate nearly 362% more $CO_2$, equivalent to approximately 3.6 times higher emissions, which remains notably high. On average across the selected urban routes for one passenger, a multicopter eVTOL produce about 2.6 times more $CO_2$ emissions compared to public transportation.

In the context of the regional use case, the comparison between lift-and-cruise eVTOLs and electric trains unequivocally demonstrate that eVTOLs emit notable more $CO_2$ than an electric train with 2 passengers.

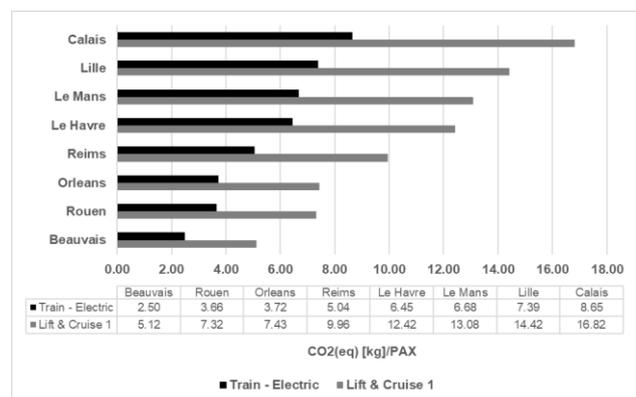

**FIG 18** Use Case: RAM – total $CO_2$ emission comparison lift-and-cruise vs. public transportation

Quantitatively, the eVTOL performs poorly with an average emissions level of approximately 197%, meaning it emits nearly 2 times more $CO_2$ than the electric train on regional routes (cf. Fig. 19).

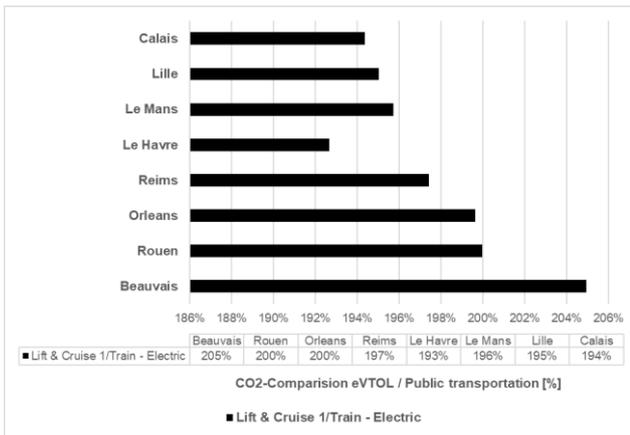

**FIG 19** Use Case: RAM – percentage $CO_2$ emission comparison lift-and-cruise vs. public transportation

This difference in $CO_2$ emissions highlights the substantial environmental advantage of electric trains over lift-and-cruise eVTOLs for regional travel. Electric trains offer a much cleaner and greener alternative, essentially reducing $CO_2$ emissions and contributing to a more sustainable transportation system.

### 4.4. $CO_2$ EMISSION EVTOL VS. HELICOPTER

The following results shows the $CO_2$ emissions per km and passenger of all investigated helicopter types and UAM vehicle configurations.

| Helicopter | $CO_2$ eq [kg/PAX-km] | eVTOL | $CO_2$ eq [kg/PAX-km] |
|---|---|---|---|
| R44 | 0.21 | Multicopter | 0.14 |
| R66 | 0.17 | Multicopter (coaxial) | 0.66 |
| H120 | 0.31 | Quadcopter | 0.15 |
| H125 | 0.27 | Lift+Cruise 1 | 0.06 |
| H135 | 0.39 | Lift+Cruise 2 | 0.10 |
| H145 | 0.35 | Vectored lift | 0.05 |
| Bell 206 | 0.37 | Tilt-rotor | 0.05 |

**TAB 9** $CO_2$ emissions per passenger seat and km flight distance

The lowest $CO_2$ emissions for any helicopter has the R66 with 0.17 kg/passenger-km. Due to the turbine engine and the higher start-up fuel consumptions, the R44 with a fuel consumption of 0.21 kg/km is used for subsequent investigations being more representative. The highest fuel consumption per passenger and flight distance has the H135 offering crucial higher redundancy, safety and flight performance compared to the R66. The highest $CO_2$ emissions for any UAM configuration type has the Quadcopter with 0.66 kg/km, the lowest the vectored thrust and the tilt-rotor concept with 0.08 kg/km each.

Considering the additional energy demand for take-off and landing the multicopter is used as representative for UAM flights in subsequent investigations being the best set-up for short flights. This type can avoid the additional transition phase required by other eVTOL types.

Fig. 20 illustrates the equivalent $CO_2$ emissions for a multicopter concept compared to the average emissions of seven selected helicopters for urban use cases. It is clear that the fuel consumption during run-up results in higher $CO_2$ emissions for the selected helicopters. For the shortest route from Place de la Concorde to the Eiffel Tower, the average emissions from the helicopters is around 0.57 kg $CO_2$ (eq), resulting in a 12% reduction in $CO_2$ emissions with the multicopter. For the longest UAM-distance to Charles de Gaulle Airport, the helicopters emit around 8.55 kg $CO_2$ (eq) per passenger on average, whereas the multicopter indicates a 51% potential reduction in $CO_2$ emissions.

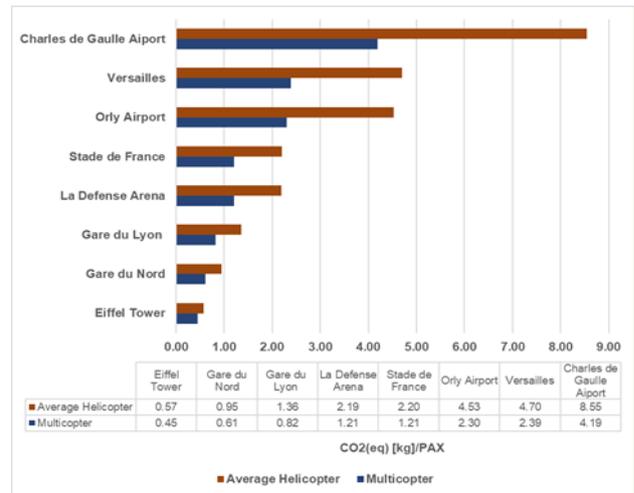

**FIG 20** $CO_2$ emissions for UAM destinations multicopter vs. average helicopter

The following figure illustrates the equivalent $CO_2$ emissions for a lift-and-cruise concept compared to the average emissions of seven selected helicopters for regional use cases. For the shortest route from Place de la Concorde to the city of Beauvais, the average emissions from the helicopters is around 20.91 kg $CO_2$ (eq), resulting in a 75% reduction in $CO_2$ emissions with the multicopter. For the longest RAM-distance to Calais, the helicopters emit around 72.38 kg $CO_2$ (eq) per passenger on average, whereas the multicopter indicates a 77% potential reduction in $CO_2$ emissions.

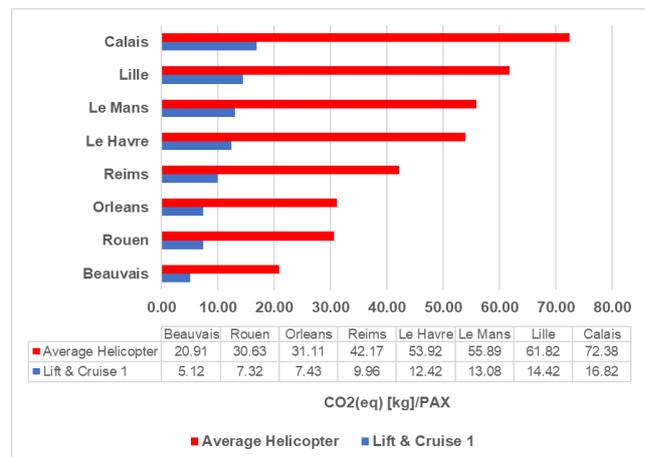

**FIG 21** $CO_2$ emissions for RAM destinations multicopter vs. average helicopter

All investigations do not consider efficiency gains of the helicopter when cabin heating or defrosting is required. This will cost additional electric energy for the UAM concept, while the helicopter can use the engine heat exchanger.

## 5. DISCUSSION

Several key elements will be discussed with respect to the sustainability of air taxis. Firstly, this paper acknowledges that while the analysis focused on time efficiency and $CO_2$ emissions of air taxis, sustainability encompasses various factors beyond these two aspects. Additional sustainable factors such as noise pollution, land usage for vertiports, or the overall infrastructure impact should be considered to provide a comprehensive evaluation of the overall sustainability of air taxis.

Secondly, in the perspective of time saving it is needed to discuss the assessment of simplistic vertiport operations and crucial factors such as boarding and de-boarding times. In real-world scenarios, the time taken for passengers to reach the vertiport from their location (door-to-Vertiport time) and the time required for boarding and de-boarding procedures can essentially impact the overall travel time. Therefore, a thorough approach, encompassing the entire travel process from the passenger's point of origin to the final destination, is essential to provide a more accurate and realistic assessment of the time-saving potential of eVTOLs in urban and regional transportation settings.

Thirdly, the paper suggests that a more in-depth comparison of $CO_2$ emissions could be achieved through a detailed analysis of the entire Product Life Cycle, which would consider emissions across all stages of an air taxi's life, from manufacturing and operation to end-of-life disposal. This comprehensive approach would provide a more detailed understanding of the environmental impact of air taxis and aid in identifying areas for improvement.

Another crucial point raised is the limited scope of the analysis, which only considered multicopters and lift-and-cruise configuration. To provide a more holistic picture, the paper suggests including various configurations, such as tilt-rotors and vectored thrust, in the calculations and comparing their performance. Different flight models may exhibit varying energy efficiencies and emissions profiles, which could influence the overall sustainability assessment.

Additionally, an essential aspect to consider is that the energy demand of eVTOLs was only calculated for a simplified mission profile (cf. Fig 5) for two scenarios UAM and RAM, respectively. To obtain a more realistic representation, further calculations should account for various operating conditions, such as hover time, and flight altitudes. Also, redundancy, flight safety and reliability where of no concern so for this investigation. The data shows that these performance indicators have an essential impact on the fuel consumption of helicopter. Another factor that was not considered is winter operations capability or all-weather capability. It is anticipated that under these conditions, the energy demand will be considerably higher, consequently leading to increased $CO_2$ emissions.

Furthermore, the paper underscores the significance of tailoring $CO_2$ emission calculations to reflect the specific energy mix of the country in which the flight routes are situated. The choice to utilize the average European electricity mix in the analysis was driven by the aim to establish a broad foundation that aligns with the overarching integration of UAM into major urban areas across Europe. This approach facilitates a comprehensive understanding of the potential sustainability impact of UAM on a pan-European scale, guiding policy decisions and frameworks for UAM adoption in diverse metropolitan contexts. However, the paper recommends a more refined approach by utilizing the distinct French electricity mix, boasting a lower $CO_2$ intensity of 58 gr/kWh, for flights conducted within France. This adaptable approach acknowledges the variance in energy sources between countries, ensuring the accuracy of sustainability data tailored to regional nuances. The dynamic nature of these energy sources also underscores the influence of contemporary political and environmental contexts.

These points emphasize the need for a broader analysis and more realistic operating conditions to obtain a comprehensive understanding of the sustainability and environmental impact of eVTOLs.

The paper brings attention to key considerations for a comprehensive evaluation of the sustainability of air taxis. It emphasizes the need to explore and include various sustainable factors beyond time efficiency and $CO_2$ emissions, conduct product-lifecycle analysis, broaden the scope of flight models, and customize emission calculations based on the specific energy mix of the region. Addressing these points would lead to a more nuanced understanding of the environmental impact of air taxis and pave the way for promoting sustainable aviation solutions.

## 6. CONCLUSION

In this paper, the timing saving and $CO_2$ emission by two eVTOL configurations for urban and regional transportation were evaluated. The analysis was compared to conventional transportation modes such as cars, public transportation and helicopters. In the realm of $CO_2$ emissions, the analysis uncovers a valuable divergence between urban and regional scenarios.

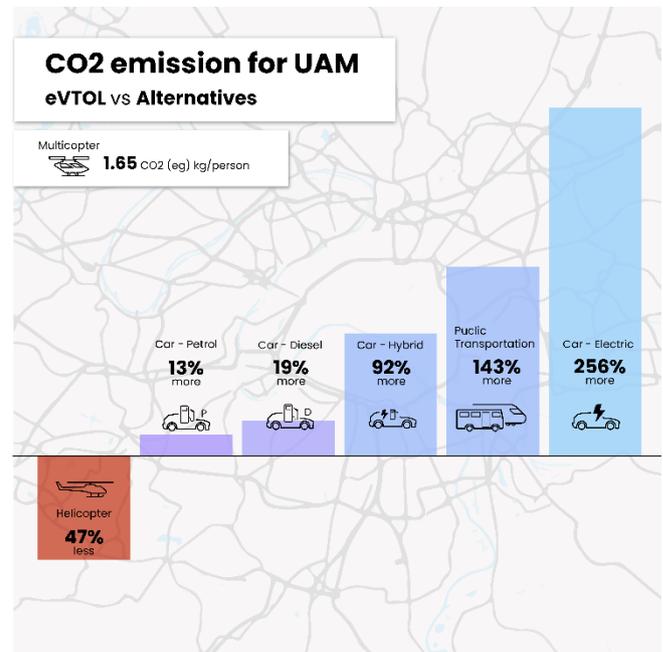

**FIG 21** $CO_2$ emissions comparison for UAM: Multicopter vs. alternatives

Urban eVTOL operations are aiming to offer a greener and more efficient alternative to conventional transportation. However, due to their elevated energy consumption, they demonstrate relatively lesser ecological friendliness compared to certain other modes of transportation. Specifically, for the UAM-use case, a Multicopter consumes on average UAM-mission 1.65 kg of $CO_2$ equivalent per person. When compared to other means of transport, the Multicopter consumes 47% less than the average conventional helicopter, making it a more environmentally conscious choice for urban air mobility. However, it does consume 13% more than a petrol car, 19% more than a diesel car, 92% more than a hybrid car, 143% more than metro-bus-tram, and 256% more than an electric car.

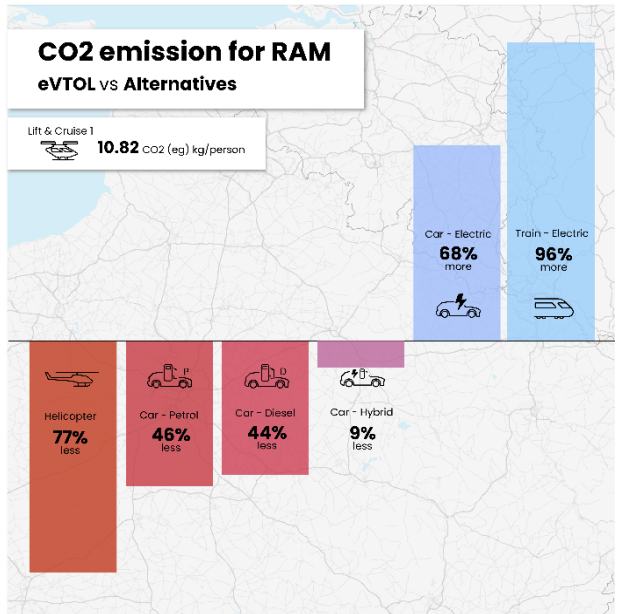

**FIG 22** $CO_2$ emissions comparison for RAM: Lift-and-cruise vs. alternatives

In a regional context, eVTOLs, such as the Lift-and-Cruise 1, showcase substantial $CO_2$ reductions compared to average conventional helicopters, petrol, diesel, and hybrid vehicles. Specifically, the Lift-and-Cruise 1 eVTOL consumes 77% less $CO_2$ equivalent than the average conventional helicopter, 46% less than a petrol car, 44% less than a diesel car, and 9% less than a hybrid car. This highlights the valuable $CO_2$ emission comparison between eVTOLs and different conventional transportation modes. However, it is crucial to note that eVTOLs emit significantly 68% more $CO_2$ than electric vehicles and 96% more than electric trains on selected regional routes.

Consequently, while the adoption of eVTOLs can significantly reduce the carbon footprint compared to using conventional helicopters and certain ground vehicles, it is essential to consider the broader spectrum of available transportation modes, as ground-based electric vehicles and public transportation still offer lower $CO_2$ emissions for urban and regional transit.

The following key findings can be highlighted:

- eVTOLs can save more than 20 min on average compared to cars and public transportation on urban level (<50 km)
- eVTOLs can save essential time on regional level (< 300 km) compared to cars around of 76 minutes and 69 minutes compared to train
- electric VTOL concepts can reduce the operational $CO_2$ emissions compared to combustion engine driven helicopter flights
- eVTOLs are more environmentally friendly than helicopter but not more than other transport options in urban areas, as they consume more energy
- eVTOLs can save $CO_2$ emissions compared to combustion engine cars on regional routes
- Electric trains are the most environmentally friendly alternative for regional transport

These results highlight the need to continue research and development to improve the environmental performance of eVTOL technology and prioritize the introduction of greener transportation alternatives to reduce the environmental impact of greenhouse gas emissions. The most obvious impacts are energy requirements during takeoff, startup, and shutdown, especially for UAM flights, and lower $CO_2$ emissions during cruise for regional flights.

As battery technology advances, eVTOLs are likely to become increasingly efficient, paving the way for a cleaner and more sustainable mode of transportation in the future. However, it is also prudent to focus on renewable energy sources. In summary, the carbon footprint of UAM vehicles is largely dependent on the type of energy source used to power them.

Efforts to promote and invest in efficient transportation and greener technologies have an important role to play in achieving a more sustainable and environmentally conscious future.

## ACKNOWLEDGEMENT

The research presented in this paper is part of the research activity on the project HorizonUAM carried out by the Department of Unmanned Aerial Systems (UAS) at the Institute of Flight Guidance by the German Aerospace Centre (DLR).

I extend my gratitude to Atul Kumar, Nicolas Brieger, Markus Engelhardt, Thuysi Dao, and Veruska Mazza Rodrigues Dias for their valuable contributions to this paper. I would also like to express my appreciation to all other participants who engaged in discussions that contributed to shaping this critical narrative, as their collective input was essential in developing a comprehensive understanding of the complexities and implications surrounding AAM and eVTOLs.

## COMPETING INTERESTS

The authors have no competing interests to declare that are relevant to the content of this article.

**APPENDIX**

The detailed calculated results for time saving on UAM level:

Use-Case:     UAM - Intra-City     (< 50 km)
Start:        Place de la Concorde (48.880079377411434, 2.3209529783314298)

| Destination: UAM | Flight range [km] | Flight time [min] | Car distance [km] | Car time [min] | Train time [min] |
|---|---|---|---|---|---|
| Eiffel Tower | 1.94 | 2 | 2,8 | 7 | 20 |
| Gare du Nord | 3.21 | 3 | 3,4 | 25 | 21 |
| Gare du Lyon | 4.59 | 4 | 5,5 | 26 | 16 |
| La Défense Arena | 7.42 | 5 | 9,4 | 35 | 24 |
| Stade de France | 7.45 | 5 | 15,9 | 35 | 34 |
| Orly Airport | 15.33 | 10 | 16,4 | 37 | 57 |
| Versailles | 15.92 | 10 | 19,4 | 38 | 49 |
| Charles de Gaulle Airport | 28.94 | 19 | 32,9 | 46 | 03 |

| Destination: UAM | Latitude and Longitude |
|---|---|
| Eiffel Tower | 48.85860133062431, 2.294521555637876 |
| Gare du Nord | 48.88135187311856, 2.355220655293813 |
| Gare du Lyon | 48.84467549552589, 2.374281457292527 |
| La Défense Arena | 48.89563213687809, 2.22961956012617 |
| Stade de France | 48.92601522501372, 2.3602917068864584 |
| Orly Airport | 48.7302053462825, 2.365637273265505 |
| Versailles | 48.80705182917126, 2.120254946559202 |
| Charles de Gaulle Airport | 49.01171630623034, 2.5518109714304003 |

The detailed calculated results for time saving on RAM level:

Use-Case:     RAM                  (< 300 km)
Start:        Place de la Concorde (48.880079377411434, 2.3209529783314298)

| Destination: RAM | Flight range [km] | Flight time [hr:min] | Car distance [km] | Car time [hr:min] | Train time [hr:min] |
|---|---|---|---|---|---|
| Calais | 240 | 01:21 | 298 | 03:13 | 04:09 |
| Le Mans | 185 | 01:03 | 213 | 02:25 | 02:35 |
| Beauvais | 69 | 00:24 | 97 | 01:21 | 01:37 |
| Lille | 205 | 01:09 | 222 | 02:31 | 02:31 |
| Orleans | 103 | 00:36 | 154 | 01:58 | 02:12 |
| Le Havre | 179 | 01:01 | 198 | 02:22 | 03:21 |
| Reims | 140 | 00:48 | 157 | 01:30 | 02:30 |
| Rouen | 101 | 00:35 | 129 | 01:47 | 02:38 |

| Destination: RAM | Latitude and Longitude |
|---|---|
| Calais | 51.014019961073494, 1.9589627300023547 |
| Le Mans | 47.971754926169446, 0.19406329470439904 |
| Beauvais | 49.47374521207076, 2.1076144873066 |
| Lille | 50.63954613506171, 3.1071012113886605 |
| Orleans | 47.940949692303505, 2.1667537797583964 |
| Le Havre | 49.56180450545073, 0.09679634369569831 |
| Reims | 49.228246054031615, 4.159641967594461 |
| Rouen | 49.40050050927563, 1.1862262267802346 |

The detailed energy demand and $CO_2$ emission for UAM:

| | eVTOL | |
|---|---|---|
| **Destination: UAM** | **Energy demand/PAX [kWh]** | **$CO_2$eq/PAX [kg]** |
| Eiffel Tower | 1.9 | 0.45 |
| Gare du Nord | 2.71 | 0.61 |
| Gare du Lyon | 3.63 | 0.82 |
| La Défense Arena | 5.37 | 1.21 |
| Stade de France | 5.38 | 1.22 |
| Orly Airport | 10.17 | 2.30 |
| Versailles | 10.58 | 2.39 |
| Charles de Gaulle Airport | 18.55 | 4.19 |

| | **Car - Petrol** | | **Car - Diesel** | | **Car - Hybrid** | | **Car - Electric** | |
|---|---|---|---|---|---|---|---|---|
| **Destination: UAM** | **Energy demand/ PAX [kWh]** | **$CO_2$eq/ PAX [kg]** | **Energy demand/ PAX [kWh]** | **$CO_2$eq/ PAX [kg]** | **Energy demand/ PAX [kWh]** | **$CO_2$eq/ PAX [kg]** | **Energy demand/ PAX [kWh]** | **$CO_2$eq/ PAX [kg]** |
| Eiffel Tower | 0.96 | 0.31 | 0.91 | 0.29 | 0.45 | 0.18 | 0.21 | 0.10 |
| Gare du Nord | 1.17 | 0.37 | 1.10 | 0.36 | 0.55 | 0.22 | 0.26 | 0.12 |
| Gare du Lyon | 1.89 | 0.61 | 1.79 | 0.58 | 1.10 | 0.36 | 0.41 | 0.19 |
| La Défense Arena | 3.24 | 1.03 | 3.05 | 0.99 | 1.53 | 0.61 | 0.71 | 0.33 |
| Stade de France | 5.48 | 1.75 | 5.17 | 1.67 | 2.58 | 1.03 | 1.19 | 0.56 |
| Orly Airport | 5.65 | 1.80 | 5.33 | 1.72 | 2.66 | 1.07 | 1.23 | 0.57 |
| Versailles | 6.68 | 2.13 | 6.30 | 2.04 | 3.15 | 1.26 | 1.46 | 0.68 |
| Charles de Gaulle Airport | 11.33 | 3.62 | 10.69 | 3.45 | 5.35 | 2.14 | 2.47 | 1.15 |

| | **Public Transport** | |
|---|---|---|
| **Destination: UAM** | **Energy demand/PAX [kWh]** | **$CO_2$eq/PAX [kg]** |
| Eiffel Tower | 0.29 | 0.12 |
| Gare du Nord | 0.48 | 0.21 |
| Gare du Lyon | 0.69 | 0.29 |
| La Défense Arena | 1.11 | 0.47 |
| Stade de France | 1.12 | 0.48 |
| Orly Airport | 2.30 | 0.98 |
| Versailles | 2.39 | 1.02 |
| Charles de Gaulle Airport | 4.34 | 1.85 |

The detailed energy demand and $CO_2$ emission for RAM:

|  | eVTOL ||
| Destination: RAM | Energy demand/PAX [kWh] | $CO_2$eq/PAX [kg] |
| --- | --- | --- |
| Beauvais | 22.68 | 5.12 |
| Rouen | 32.41 | 7.32 |
| Orleans | 32.86 | 7.43 |
| Reims | 44.05 | 9.96 |
| Le Havre | 54.97 | 12.42 |
| Le Mans | 57.88 | 13.08 |
| Lille | 63.79 | 14.42 |
| Calais | 74.44 | 16.82 |

|  | Car - Petrol || Car - Diesel || Car - Hybrid || Car - Electric ||
| Destination: RAM | Energy demand/ PAX [kWh] | $CO_2$eq/ PAX [kg] | Energy demand/ PAX [kWh] | $CO_2$eq/ PAX [kg] | Energy demand/ PAX [kWh] | $CO_2$eq/ PAX [kg] | Energy demand/ PAX [kWh] | $CO_2$eq/ PAX [kg] |
| --- | --- | --- | --- | --- | --- | --- | --- | --- |
| Beauvais | 33.41 | 10.67 | 31.52 | 10.19 | 15.76 | 6.31 | 7.28 | 3.40 |
| Rouen | 44.43 | 14.19 | 41.92 | 13.55 | 20.96 | 8.39 | 9.68 | 4.52 |
| Orleans | 53.04 | 16.94 | 50.04 | 16.17 | 25.02 | 10.01 | 11.55 | 5.39 |
| Reims | 54.07 | 17.27 | 51.02 | 16.49 | 25.51 | 10.21 | 11.78 | 5.50 |
| Le Havre | 68.19 | 21.78 | 64.34 | 20.79 | 32.17 | 12.87 | 14.85 | 6.93 |
| Le Mans | 73.36 | 23.43 | 69.21 | 22.37 | 34.61 | 13.85 | 15.98 | 7.46 |
| Lille | 76.46 | 24.42 | 72.14 | 23.31 | 36.07 | 14.43 | 16.65 | 7.77 |
| Calais | 102.63 | 32.78 | 96.84 | 31.29 | 59.59 | 19.37 | 22.35 | 10.43 |

|  | Electric train ||
| Destination: UAM | Energy demand/PAX [kWh] | $CO_2$eq/PAX [kg] |
| --- | --- | --- |
| Beauvais | 10.42 | 2.50 |
| Rouen | 15.26 | 3.66 |
| Orleans | 15.50 | 3.72 |
| Reims | 21.01 | 5.04 |
| Le Havre | 26.86 | 6.45 |
| Le Mans | 27.84 | 6.68 |
| Lille | 30.80 | 7.39 |
| Calais | 36.06 | 8.65 |